\documentclass[twocolumn,showpacs,preprintnumbers,amsmath,amssymb]{revtex4}

\usepackage{graphicx}
\usepackage{dcolumn}
\usepackage{bm}

\newtheorem{Thm}{Theorem}
\newtheorem{Lem}[Thm]{Lemma}
\newtheorem{Cor}[Thm]{Corollary}

\begin{document}

\preprint{Th-DMFT-2}

\title{A uniqueness theorem in a density-matrix functional theory}

\author{Koichi Kusakabe}
\email{kabe@mp.es.osaka-u.ac.jp}
\affiliation{%
Graduate School of Engineering Science, Osaka University\\
1-3 Machikaneyama-cho, Toyonaka, Osaka 560-8531, Japan
}%

\date{\today}

\begin{abstract}
Uniqueness of effective interaction defined in 
an extension of the Kohn-Sham theory is proved, 
if the model with a non-degenerate ground state exists and to reproduce 
a correlation function as well as the single-particle density 
of an electron system. 
The two-body interaction term is regarded as a Hubbard-type 
short range interaction term. 
The interaction strength is a functional 
of an element of a two-body reduced density matrix 
of the electron system. 
The uniqueness theorem gives a basic principle 
for effective description of electron systems. 
\end{abstract}

\maketitle


Progress in theoretical description of many-electron systems 
and the high-performance-computational techniques have allowed us to 
predict new functionality of materials and even to design unknown materials 
by theoretical model calculations. 
The density-functional theory 
(DFT)\cite{Hohenberg-Kohn,Kohn-Sham,Levy79,Levy82,Lieb83} 
has played a central role to realize reliable and powerful skills 
for the material scientists. 
Moreover, DFT is valuable for pure physics since it provides a way to 
create an effective Hamiltonian of an interacting quantum system. 

By introducing an idea that the single-particle density is 
reproduced by an effective many-body system, Kusakabe derived 
a kind of extension of the Kohn-Sham theory.\cite{Kusakabe01} 
In our extended Kohn-Sham scheme, we have an effective many-body 
Hamiltonian expressed in second quantization. 
The creation and annihilation operators 
for artificial particles are defined using single-particle 
wave functions which may be expanded in Kohn-Sham orbitals. 
The effective potential for our modified Kohn-Sham single-particle 
equation is defined with a residual exchange-correlation energy 
functional $E_{\rm rxc}[n]$. 
Here $n({\bf r})$ is the electron density. 
This residual term is determined via selection of the 
interaction terms introduced explicitly in the model. 
If an interaction term is written in terms of a localized orbital, 
the interaction may be interpreted 
as the Hubbard-type short-range interaction. 
Shift in interaction strength may change 
$E_{\rm rxc}[n]$, but the formulation holds without uncertainty. 
This arbitrariness means that 
our effective theory is not uniquely determined 
within the framework of the density-functional theory. 

This problem of uncertainty is solved in this paper 
by formulating the theory as a density-matrix-functional theory (DMFT). 
DMFT is originally written as an extension of DFT. 
In the usual DMFT, $m$-body reduced density matrices 
$\gamma_m({\bf r}_1',\cdots,{\bf r}_m';{\bf r}_1,\cdots,{\bf r}_m)$ are used 
as the basic variable.\cite{DMFT_review,Donnelly-Parr} 
Here ${\bf r}_i$ is a space coordinate for electrons. 
Since a two-body reduced density matrix gives 
the exact interaction energy, knowledge of the reduced density matrices 
determines the true ground state energy of a many-electron system. 
However, there remains the $N$-representability problem in the known DMFT. 

On the other hand, a natural form of two-body reduced density matrices 
appears in our extended Kohn-Sham scheme.\cite{Kusakabe01} 
The form representing density fluctuation can be 
a density-density correlation function in a localized orbital. 
We will show that the simplest form is positive real-valued and 
its range is from 0 to 1. 
Our model gives a minimization process of a wave-function functional 
and the minimum of the model is shown to exist. 
If we request that the fluctuation in the original Coulomb system 
is reproduced in our model, which should have a non-degenerate ground state, 
the model of the many-electron system is uniquely determined. 
In a specified situation, 
we will conclude an existence condition of the properly defined 
extended Kohn-Sham model with an interaction parameter $U$ as 
a density-matrix functional. 

We may choose relevant orbitals and 
introduce an operator detecting fluctuation per each orbital. 
To simplify the discussion, we consider at first an impurity problem 
with a single relevant orbital $\phi_i({\bf r})$ 
at an atomic center in a bulk. 
This situation naturally gives a Kondo problem. 
However, it is only technical to consider an impurity problem. 
Once our strategy is given for this typical case, 
further possible extension becomes trivial. 

We introduce an operator 
defining a correlation function 
$\langle\underline{n}_i^2\rangle$, which is 
the density-density correlation function on $\phi_i({\bf r})$. 
\begin{equation}
\langle \underline{n}_i^2 \rangle 
\equiv
\langle 
(n_{i,\uparrow}+n_{i,\downarrow}-\bar{n}_{i,\uparrow}-\bar{n}_{i,\uparrow})^2
\rangle
\end{equation}
Here, $\phi_i({\bf r})$ is normalized and 
defines associated creation and annihilation operators 
denoted by $c_{i,\sigma}^\dagger$ and $c_{i,\sigma}$. 
The number operator $n_{i,\sigma}=c_{i,\sigma}^\dagger c_{i,\sigma}$ 
is used. Expectation values 
$\bar{n}_{i,\sigma}\equiv \langle\Psi |n_{i,\sigma}|\Psi\rangle$ 
are taken for a many-body state $|\Psi\rangle$, 
which is a trial state unless explicitly specified. 

The density $n({\bf r})$ and the expectation value 
$\langle \underline{n}_i^2 \rangle$ given by $|\Psi\rangle$ are 
the fundamental variables in our theory. 
For the ground state $|\Phi_G\rangle$ of the Coulomb system, 
$n_G({\bf r})$, $\langle \underline{n}_i^2 \rangle_G$ 
given by $|\Phi_G\rangle$ 
will be used. 
We are seeking for a model in which 
$n_G({\bf r})$ and $\langle \underline{n}_i^2 \rangle_G$ 
are reproduced exactly. 

We first show the next lemma. 
\begin{Lem}
\label{Lemmma-region}
\item $\langle \underline{n}_i^2 \rangle$ is real. 
The next inequality holds. 
\begin{equation}
0 \le \langle\underline{n}_i^2\rangle \le 1 \; .
\end{equation}
\end{Lem}
Proof: Since $n_{i,\sigma}$ is Hermitian, 
$\langle\underline{n}_i^2\rangle$ is real and positive-semi-definite. 
Arbitrary $|\Psi\rangle$ is expanded as, 
\[|\Psi\rangle=A_0 |\Psi_0\rangle
+A_1 c_{i,\uparrow}^\dagger|\Psi_1\rangle
+A_2 c_{i,\downarrow}^\dagger|\Psi_2\rangle
+A_3 c_{i,\uparrow}^\dagger c_{i,\downarrow}^\dagger|\Psi_3\rangle\; .\]
Here each $|\Psi_i\rangle$ ($i=0,1,2,3$) satisfies 
$c_{i,\sigma}|\Psi_i\rangle=0$. 
Coefficients $A_i$ satisfies a normalization condition 
$|A_0|^2+|A_1|^2+|A_2|^2+|A_3|^2=1$. 
Using $A_i$, we have 
\begin{eqnarray}
\lefteqn{\langle\underline{n}_i^2\rangle} \nonumber \\
&=&2|A_3|^2+\left( |A_1|^2+|A_2|^2+2|A_3|^2\right) \nonumber \\
&\times&\left\{ 1- \left( |A_1|^2+|A_2|^2+2|A_3|^2\right)\right\} \; .
\end{eqnarray}
This expression has the maximum value 1, only if 
$|A_0|=|A_3|=1/\sqrt{2}$ and $|A_1|=|A_2|=0$. 

We require $\langle \underline{n}_i^2 \rangle$ 
to be reproduced in our model system. 
There are some possible manners 
to formulate the density matrix functional theory. 
One might consider generalization of 
constrained search of Levy\cite{Higuchi-Higuchi} in order to make 
a density matrix functional theory. However, this is not the only way. 
We utilize Levy's universal energy functional $F[n]$ given by 
\[ 
F[n]
=\min_{\Psi'\rightarrow n}
\langle \Psi' | \hat{T}+\hat{V}_{\rm ee} | \Psi' \rangle \; , 
\]
to formulate our theory. 
Using the electron field operators $\psi^\dagger_{\sigma}({\bf r})$ and 
$\psi_{\sigma}({\bf r})$, the kinetic energy operator 
$\hat{T}$ and the Coulomb interaction $\hat{V}_{\rm ee}$ 
are given by, 
\[\hat{T}=-\frac{\hbar^2}{2m} \int \! d^3r\, \sum_{\sigma} 
\lim_{{\bf r}' \rightarrow {\bf r}} 
\psi^\dagger_{\sigma}({\bf r}') \Delta_{\bf r} \psi_{\sigma}({\bf r}) \; .
\]
\[
\hat{V}_{\rm ee} = \frac{1}{2} \int \! d^3r \, d^3r' \,
\frac{e^2}{|{\bf r}-{\bf r}'|} \sum_{\sigma,\sigma'}
\psi^\dagger_{\sigma}({\bf r}) \psi^\dagger_{\sigma'}({\bf r}') 
\psi_{\sigma'}({\bf r}') \psi_{\sigma}({\bf r}) \; .
\]
It is known that a minimizing $\Psi'$ in the definition of $F[n]$ exists, 
although the convexity of $F[n]$ is denied.\cite{Lieb83} 

Now we consider the reduced interaction term, which may be regarded 
as correlation or fluctuation given by, 
\begin{equation}
\langle \Psi | V_{\rm red} | \Psi \rangle 
=\frac{U}{2}\langle \Psi |\underline{n}_i^2 |\Psi \rangle \; .
\end{equation}
We define a functional $\bar{G}[\Psi]$, 
which is an extension of Hadjisavvas-Theophilou's 
functional.\cite{Hadjisavvas-Theophilou,Kusakabe01} 
\begin{eqnarray}
\label{ex-Kohn-Sham}
\bar{G}[\Psi]&=&
\langle \Psi | \hat{T}+\hat{V}_{\rm red} | \Psi \rangle
-\min_{\Psi'\rightarrow n_\Psi}
\langle \Psi' | \hat{T}+\hat{V}_{\rm red} | \Psi' \rangle \nonumber \\
&+&F[n_\Psi]
+\int d^3r v_{\rm ext}({\bf r}) n_\Psi ({\bf r}) \nonumber \\
&=&
\langle \Psi | \hat{T}+\hat{V}_{\rm red} | \Psi \rangle
+\frac{1}{2}\int\frac{n_\Psi({\bf r})n_\Psi({\bf r}')}
{|{\bf r}-{\bf r}'|}d^3rd^3r' 
\nonumber \\
&&+E_{\rm rxc}[n_\Psi] 
+\int d^3r v_{\rm ext}({\bf r}) n_\Psi ({\bf r}) \; . 
\end{eqnarray}
Here, the residual-exchange-correlation functional is, 
\begin{eqnarray}
\lefteqn{E_{\rm rxc}[n_\Psi]} \nonumber \\
&=& \min_{\Psi'\rightarrow n_\Psi}
\langle \Psi' | \hat{T}+\hat{V}_{\rm ee} | \Psi' \rangle 
-\min_{\Psi'\rightarrow n_\Psi}
\langle \Psi' | \hat{T}+\hat{V}_{\rm red}| \Psi' \rangle \nonumber \\
&-&\frac{1}{2}\int\frac{n_\Psi({\bf r})n_\Psi({\bf r}')}
{|{\bf r}-{\bf r}'|}d^3rd^3r' \; . \nonumber 
\end{eqnarray}
In general, the definition of $V_{\rm red}$ determines 
$E_{\rm rxc}$. Varying $U$ may change $E_{\rm rxc}$. 
Any $U$ is allowed to determine a model. However, 
almost all of the models are not proper, since 
fluctuation may not be reproduced. 
To make $U$ dependence explicit, 
we utilize a symbol $\bar{G}_U[\Psi]$ below. 

We next prove a set of exact statements in our theory. 
\begin{Lem}
\label{Lemma-statements}
Assume that the ground state of a Coulomb system 
given by $v_{\rm ext}({\bf r})$ is non-degenerate. 
{\rm i)} The ground state $\Psi$ of a corresponding extended-Kohn-Sham model 
$\bar{G}_U[\Psi]$ with a given positive $U$ exists. 
{\rm ii)} For fixed $n({\bf r})$, 
$\bar{F}(U)=\min_{\Psi\rightarrow n}\langle\Psi|\hat{T}+\frac{U}{2}
\underline{n}_i^2|\Psi\rangle$ is a continuous function of $U$. 
{\rm iii)} If a state $|\Psi\rangle$ is the ground state 
of $\bar{G}_{U_1}[\Psi]$ and $\bar{G}_{U_2}[\Psi]$ with 
$0\le U_1 < U_2$ simultaneously, 
$|\Psi\rangle$ is the ground state of $\bar{G}_{U}[\Psi]$ 
in a finite range $[U_1,U_2]$ of $U$. 
\end{Lem}
Proof: 
The ground state, which gives the minimum of $\bar{G}_U[\Psi]$, 
satisfies two conditions. 
(1) $|\Psi\rangle \rightarrow n_G({\bf r})$. 
(2) $|\Psi\rangle$ minimizes $\langle \Psi | \hat{T}+\frac{U}{2}
\underline{n}_i^2 | \Psi \rangle $. 
The first statement i) is proved following the 
same argument in the proof of the theorem 3.3 in \cite{Lieb83}. 
The only requirement which has to be shown is 
that $\langle \hat{T}+\frac{U}{2}
\underline{n}_i^2 \rangle $ is
a positive quadratic form, which is trivial from the definition. 

The second statement ii) is now discussed. 
Assume that $\bar{F}(U)$ is not continuous when $U=U_0>0$. 
This is equivalent to a statement that 
$^\exists \varepsilon >0, \; ^\forall \delta>0, \; 
^\exists U>0, \; s.t. \; 
\{ |U-U_0|<\delta \; \& \; |\bar{F}(U)-\bar{F}(U_0)|>\varepsilon \}$.
For simplicity, let us further assume that $0<U-U_0<\delta$ and 
$\bar{F}(U)-\bar{F}(U_0)>\varepsilon$. 
If we let $\delta=\varepsilon$, we have, 
\begin{eqnarray}
\lefteqn{
\min_{\Psi'\rightarrow n} 
\langle \Psi' | \hat{T}+ \frac{U}{2}\underline{n}_i^2| \Psi' \rangle } 
\nonumber \\
&>&
\min_{\Psi'\rightarrow n} \langle \Psi' | \hat{T}+ \frac{U_0}{2}\underline{n}_i^2| \Psi' \rangle + \varepsilon
\nonumber \\
&=&
\langle \Psi_0 | \hat{T}+ \frac{U_0}{2}\underline{n}_i^2| \Psi_0 \rangle 
+ \varepsilon 
\nonumber \\
&=&
\langle \Psi_0 | \hat{T}+ \frac{U}{2}\underline{n}_i^2| \Psi_0 \rangle 
+\frac{U_0-U}{2} 
\langle \Psi_0 | \underline{n}_i^2| \Psi_0 \rangle 
+ \varepsilon 
\nonumber \\
&>&
\langle \Psi_0 | \hat{T}+ \frac{U}{2}\underline{n}_i^2| \Psi_0 \rangle 
- \frac{\delta}{2} + \varepsilon 
\nonumber \\
&=&
\langle \Psi_0 | \hat{T}+ \frac{U}{2}\underline{n}_i^2| \Psi_0 \rangle 
+ \frac{\varepsilon}{2} \; . \nonumber 
\end{eqnarray}
Here we have utilized the lemma \ref{Lemmma-region}. 
This inequality contradicts the definition of 
$\min_{\Psi'\rightarrow n} \langle \Psi' | \hat{T}+ \frac{U}{2}\underline{n}_i^2| \Psi' \rangle $. 
In other possible cases, we have the same conclusion. 

The third statement iii) is shown as follows. 
We define 
${\cal H}_G=\{|\Psi'\rangle||\Psi'\rangle \rightarrow n_G({\bf r})\}$. 
Assume that a $|\Psi\rangle$ minimizes 
both $\bar{G}_{U_1}[\Psi]$ and $\bar{G}_{U_2}[\Psi]$. 
Then for any $|\Psi'\rangle \in {\cal H}_G$ we have the next inequality 
with two parameters $\alpha\ge 0$ and $\beta\ge 0$, 
which do not simultaneously become zero. 
\begin{eqnarray}
\lefteqn{\alpha \langle\Psi'|\hat{T}+\frac{U_1}{2}\underline{n}_i^2|\Psi'\rangle
+\beta \langle\Psi'|\hat{T}+\frac{U_2}{2}\underline{n}_i^2|\Psi'\rangle}
\nonumber \\
&\ge &
\alpha \langle\Psi|\hat{T}+\frac{U_1}{2}\underline{n}_i^2|\Psi\rangle
+\beta \langle\Psi|\hat{T}+\frac{U_2}{2}\underline{n}_i^2|\Psi\rangle \; .
\nonumber 
\end{eqnarray}
This inequality ensures 
that for $U'=\frac{U_1\alpha+U_2\beta}{2(\alpha+\beta)}$ 
\[
\langle\Psi'|\hat{T}+\frac{U'}{2}\underline{n}_i^2|\Psi'\rangle
\ge 
\langle\Psi|\hat{T}+\frac{U'}{2}\underline{n}_i^2|\Psi\rangle
\; .
\]
A minimizing $|\Psi'\rangle$ of $\bar{G}_{U'}[\Psi]$
is given by $|\Psi\rangle$. 
Thus $|\Psi\rangle$ has to be always a ground state of 
$\bar{G}_U[\Psi]$, if $U$ satisfies $U_1 \le U \le U_2$. 

We now define a concept of ``a proper model''. 
In the present context, a model is properly determined, 
if its lowest level is non-degenerate and if 
it reproduces $n_G({\bf r} )$ and 
$\langle\underline{n}_i^2\rangle_G$ at the same time. 
The second condition should be changed, if the definition 
of $V_{\rm red}$ is changed. 
The central problem of the present study is 
summarized in the next question. 
If we find a proper model $\bar{G}_U[\Psi]$ by optimizing $U$, 
is there another $U'$ which do give a proper model? 
Our uniqueness theorem given below denies 
existence of the other model. 

Suppose we found a proper model with $U=U_1$. 
The ground state $|\Psi\rangle$ of $\bar{G}_{U_1}[\Psi]$ 
is non-degenerate and thus unique. 
If we shift $U$, $|\Psi\rangle$ may found as a ground state 
of $\bar{G}_{U_2}[\Psi]$ with {\it e.g.} $U_2>U_1$, 
in which the lowest level may be degenerate. 
Lemma \ref{Lemma-statements} iii) tells us that, 
in a finite region $[U_1,U_2]$, $|\Psi\rangle$ is always 
the ground state of the model $\bar{G}_{U}[\Psi]$. 
The ground state $|\Psi\rangle$ of the model does not feel 
the short range interaction by $U$. 
We call the situation ``the irrelevant $U$ case''
This happens when $\phi_i$ is out of the system 
with no electron occupying $\phi_i$ 
or $\phi_i$ is occupied by a single electron 
without hybridization. 
Another case is when the system shows the complete ferromagnetism. 
We do not intend to consider these situations. 
In real materials, we have finite hybridization matrix elements 
between $\phi_i$ and a neighboring orbital. 
Without the large magnetic field, 
the true complete ferromagnetism never exists in real solids. 
Thus we assume that $|\Psi\rangle$ does not 
appear as a ground state of $\bar{G}_{U}[\Psi]$ except for $U=U_1$. 
We do not assume, however, that $\bar{G}_{U}[\Psi]$ may have 
degeneracy at $U\neq U_1$. 
We now state our main theorem. 

\begin{Thm}
\label{Theorem-Uniqueness}
Assume that the ground state of a Coulomb system is non-degenerate. 
A proper extended-Kohn-Sham model given by $\bar{G}_{U}[\Psi]$ 
which has a non-degenerate ground state and reproduces both 
$n_G({\bf r})$ and $\langle \underline{n}_i^2 \rangle_G$ 
is uniquely determined, or it does not exist. 
\end{Thm}
Proof: 
Assume that we have $U_1\neq U_2$ and two 
extended Kohn-Sham models, {\it i.e.} 
$\bar{G}_{U_1}[\Psi]$ and $\bar{G}_{U_2}[\Psi]$, 
reproduce $n_G({\bf r})$ and $\langle \underline{n}_i^2 \rangle_G$. 
Because we do not consider an irrelevant $U$ case, 
$|\Psi_{U_1}\rangle$ and $|\Psi_{U_2}\rangle$ are different with each other. 
Let us define $E_{\rm GS}^{(i)}$ be the ground-state energy 
and $E_{\rm rxc}^{(i)}[n_G]$ be 
the residual exchange-correlation energy 
of these models. ($i=1,2$.) 
We have the next inequality. 
\begin{eqnarray}
\lefteqn{E_{\rm GS}^{(2)}}\nonumber \\
&\equiv&
\langle\Psi_{U_2}|\hat{T}+\frac{U_2}{2}\underline{n}_i^2|\Psi_{U_2}\rangle
+E_{\rm rxc}^{(2)}[n_G] \nonumber \\
&+&\frac{e^2}{2}\int d^3r d^3r' 
\frac{n_G({\bf r})n_G({\bf r}')}{|{\bf r}-{\bf r}'|} 
+\int d^3r v_{\rm ext}({\bf r} )n_G({\bf r}) \nonumber \\
&=&
\langle\Psi_{U_2}|\hat{T}+\frac{U_1}{2}\underline{n}_i^2|\Psi_{U_2}\rangle
+E_{\rm rxc}^{(1)}[n_G] \nonumber \\
&+&\frac{e^2}{2}\int d^3r d^3r' 
\frac{n_G({\bf r})n_G({\bf r}')}{|{\bf r}-{\bf r}'|}
+\int d^3r v_{\rm ext}({\bf r} )n_G({\bf r}) \nonumber \\
&+&\frac{1}{2}(U_2-U_1)\langle\Psi_{U_2}|\underline{n}_i^2|\Psi_{U_2}\rangle
+E_{\rm rxc}^{(2)}[n_G]-E_{\rm rxc}^{(1)}[n_G] \nonumber \\
&>&
\langle\Psi_{U_1}|\hat{T}+\frac{U_1}{2}\underline{n}_i^2|\Psi_{U_1}\rangle
+E_{\rm rxc}^{(1)}[n_G]\nonumber \\
&+&\frac{e^2}{2}\int d^3r d^3r' 
\frac{n_G({\bf r})n_G({\bf r}')}{|{\bf r}-{\bf r}'|} 
+\int d^3r v_{\rm ext}({\bf r} )n_G({\bf r}) \nonumber \\
&+&\frac{1}{2}(U_2-U_1)\langle\Psi_{U_1}|\underline{n}_i^2|\Psi_{U_1}\rangle
+E_{\rm rxc}^{(2)}[n_G]-E_{\rm rxc}^{(1)}[n_G] \nonumber \\
&=&
\langle\Psi_{U_1}|\hat{T}+\frac{U_2}{2}\underline{n}_i^2|\Psi_{U_1}\rangle
+E_{\rm rxc}^{(2)}[n_G] \nonumber \\
&+&\frac{e^2}{2}\int d^3r d^3r' 
\frac{n_G({\bf r})n_G({\bf r}')}{|{\bf r}-{\bf r}'|}
+\int d^3r v_{\rm ext}({\bf r} )n_G({\bf r}) \; . \nonumber
\end{eqnarray}
This contradicts that $|\Psi_{U_2}\rangle$ is the 
ground state of a model $\bar{G}_{U_2}[\Psi]$. 
The above argument does not deny non-existence of such a model. 

Thus, we have shown that 
$U$ is uniquely determined, if a proper model exists. 
Does $U$ really exist? 
If we have a case (called a case II below) 
in which $\bar{G}_{U}[\Psi]$ has always 
a unique ground state irrespective of $U$, 
we can show a sufficient condition for existence of the proper model. 
We have a next corollary. 
\begin{Thm}
\label{Theorem-forward-map}
Assume that the ground state of a Coulomb system is non-degenerate 
and that we have a case II. 
An expectation value $\langle\underline{n}_i^2\rangle$ 
given by the minimizing $|\Psi\rangle$ of $\bar{G}_{U}[\Psi]$ 
is a function of $U$. The map from $[0,\infty]$ to 
$[0,1]$ is one-to-one, continuous and thus monotone.
\end{Thm}
Proof: 
Since we have Lemma \ref{Lemmma-region}, 
we have to show that the map is a one-to-one mapping at first. 
Let us assume that 
we have two $U_i$s ($i=1,2$), satisfying $0\le U_1<U_2<\infty$, 
for which the minimizing $|\Psi_{U_i}\rangle$ of $\bar{G}_{U_i}[\Psi]$ 
reproduces $n_G({\bf r} )$ and expectation values 
$\langle\underline{n}_i^2\rangle$ at the same time. 
Using the same inequality used in the proof of Theorem \ref{Theorem-Uniqueness} 
we see that the first assumption is in contradiction to the assumption 
of the case II. 

On the continuity, let us assume the contrary. 
Then we have at least one $U_0$ on the $U$ axis, where 
the discontinuity occurs. In another word, 
\begin{eqnarray}
&&^\exists \varepsilon >0, \; ^\exists U_i>0 \; (i=1,2,\cdots\infty),
\; s.t. \; 
\{ \lim_{i\rightarrow\infty} U_i=U_0 \nonumber \\ 
&& \; \& \; 
|\langle\Psi_{U_i}|\underline{n}_i^2|\Psi_{U_i}\rangle
-\langle\Psi_{U_0}|\underline{n}_i^2|\Psi_{U_0}\rangle|>\varepsilon \} \, .
\nonumber \end{eqnarray}
This condition means that $|\Psi_{U_i}\rangle\neq |\Psi_{U_0}\rangle$. 
Now, the continuity of $\bar{F}(U)$ shown by 
Lemma 2 ii) and the selection axiom ensures that 
\[^\forall \varepsilon >0, \; 
^\exists i, \; s.t.  |\bar{F}(U_i) - \bar{F}(U_0)| < \varepsilon \; .\]
Namely, $|\Psi_{U_i}\rangle$ 
is also the ground state of $\bar{G}(U_0)$. 
However, since $|\Psi_{U_i}\rangle\neq |\Psi_{U_0}\rangle$ for all $U_i$, 
this leads to a contradiction to the uniqueness of 
$|\Psi_{U_0}\rangle$ assumed in the case II. 
Immediately, monotonous behavior is concluded. 

Now we define a set of numbers 
$R_{\phi_i}=\{r|r \in[0,1], ^\exists U \, s.t. \, 
r=\langle\Psi_U|\underline{n}_i^2|\Psi_U\rangle \}$. 
We have immediately a next corollary 
due to Theorem \ref{Theorem-forward-map}. 
\begin{Cor}
\label{Corollary-inverse-map}
There exists a one-to-one onto-mapping from 
$R_{\phi_i}$ onto $[0,\infty]$, which is 
defined by an inverse map of the function of $U$ 
in Theorem \ref{Theorem-forward-map}. 
\end{Cor}

We define a way of modelling of the Coulomb system as 
a process to find $\bar{G}_{U}[\Psi]$ 
by determining $\phi_i$ and $U$. 
We have immediately a next physical theorem. 
\begin{Cor}
\label{Corollary-Existence}
If and only if $\langle\underline{n}_i^2\rangle_G$ is in $R_{\phi_i}$, 
a proper extended Kohn-Sham model exists. 
\end{Cor}
Proof: 
The statement is trivial due to Theorem \ref{Corollary-inverse-map}. 

Here we define three minimizing states, 
$|\Psi_0\rangle$, 
$|\Psi_U\rangle$ and 
$|\Psi_C\rangle$, given by 
\begin{eqnarray}
\min_{\Psi_0\rightarrow n_\Psi} &=&
\langle\Psi_0|\hat{T}|\Psi_0\rangle \; , \nonumber \\
\min_{\Psi_U\rightarrow n_\Psi} &=&
\langle\Psi_U|\hat{T}+\frac{U}{2}\underline{n}_i^2|\Psi_U\rangle \; , \nonumber \\
\min_{\Psi_C\rightarrow n_\Psi} &=&
\langle\Psi_C|\hat{T}+\hat{V}_{\rm ee}|\Psi_C\rangle \; .\nonumber 
\end{eqnarray}
Apparently, $\Psi_C\equiv\Psi_{\rm GS}$. 
We have 
$\langle\underline{n}_i^2\rangle_{0}\equiv 
\langle\Psi_0|\underline{n}_i^2|\Psi_0\rangle$, 
$\langle\underline{n}_i^2\rangle_{U}\equiv 
\langle\Psi_U|\underline{n}_i^2|\Psi_U\rangle$, 
$\langle\underline{n}_i^2\rangle_{C}\equiv 
\langle\Psi_{\rm GS}|\underline{n}_i^2|\Psi_{\rm GS}\rangle$. 
These definitions and statements above conclude the followings. 
\begin{Lem} 
\label{Condition-on-fluc}
The following holds. 
\begin{enumerate}
\item $\langle\underline{n}_i^2\rangle_{U=\infty}=
\min_{\Psi\rightarrow n_\Psi} \langle\underline{n}_i^2\rangle$. 
\item $\langle\underline{n}_i^2\rangle_{\rm GS}\ge 
\langle\underline{n}_i^2\rangle_{U=\infty}$.
\end{enumerate}
\end{Lem}

Due to Theorem \ref{Theorem-forward-map} 
we can say the following condition for $U$ to exist. 
\begin{Thm} 
\label{Theorem-Existence}
Consider a case II. Then, 
$R_{\phi_i}=[\langle\underline{n}_i^2\rangle_{U=\infty},\langle\underline{n}_i^2\rangle_{0}]$. 
If $\langle\underline{n}_i^2\rangle_{C} < 
\langle\underline{n}_i^2\rangle_{0}$, the proper extended Kohn-Sham model 
exists and the unique $U$ given by Theorem 3 exists. 
\end{Thm}

Now we discuss several remaining problems. 
We may search for $\phi_i$ to find a proper model. 
However, we should note that both $\langle\underline{n}_i^2\rangle_G$ 
and $R_{\phi_i}$ shift, if $\phi_i$ is changed. 
For a lattice, we may consider 
\begin{equation}
\langle V_{\rm red} \rangle= \frac{U}{2} \sum_{i} 
\langle \underline{n}_i^2 \rangle \; .
\end{equation}
The same argument as the above is possible. 
Then one would find $U[\langle\underline{n}_j^2\rangle_G]$. 

To determine $\phi_i$, we may see the value of 
$\langle\underline{n}_i^2\rangle_G$. 
Practically, we can check orbital energy $\varepsilon_i$ of $\phi_i$, 
which is an expectation value of the one-body part of 
the extended Kohn-Sham equations.\cite{Kusakabe01} 
We have a Fermi level $E_F$ tentatively 
at the step to obtain the Kohn-Sham orbitals. 
If $\varepsilon_i$ is far below $E_F$ or far above $E_F$, 
the occupation of $\phi_i$ becomes 2 or 0. 
Then, we inevitably have $\langle\underline{n}_i^2\rangle_G=0$. 
If $\varepsilon_i$ is around $E_F$ and if the system is in 
a delocalized scheme, $|A_0|=|A_1|=|A_2|=|A_3|\simeq 1/2$ 
and $\langle\underline{n}_i^2\rangle_G\simeq 1/2$. 
However, if the Coulomb correlation make reduction in 
$\langle\underline{n}_i^2\rangle_G$ for 
$\varepsilon_i$ is around $E_F$, 
we are required to introduce a $U$-term to reproduce 
the density-density correlation function. 
This guide line can be a determination rule of 
minimal numbers of localized orbitals $\phi_i$, 
for which we need to set $U$-terms up. 
If singly occupied configuration 
may not appear in the true ground state, then, $|A_1|=|A_2|\simeq0$ 
and $\langle\underline{n}_i^2\rangle_G\simeq 1$. 
This condition may be given if the state is superconducting. 
In this situation, negative $U$ may be required, 
although we cannot utilize Lemma \ref{Lemma-statements} i) for this case. 

Lemma \ref{Lemma-statements} iii) leads us to four possible scenarios 
for $\bar{G}_U[\Psi]$ in the whole range of $0\le U < \infty$. 
Here we do not consider the trivial degeneracy 
coming from SU(2) symmetry of the total spin and take 
the highest weight state as a representative state. 
\begin{description}
\item[Case I]
A single state $|\Psi\rangle$ 
is always the ground state of $\bar{G}_U[\Psi]$. 
\item[Case II]
For each $U$, a unique state $|\Psi\rangle$ different from 
others appears as the ground state of $\bar{G}_U[\Psi]$. 
\item[Case III]
In a finite region $[U_1,U_2]$ of $U$, 
a state $|\Psi\rangle$ is a ground state 
of $\bar{G}_U[\Psi]$. For any other $U$, 
the state $|\Psi\rangle$ never becomes the ground state. 
\item[Case IV]
For any $U_0$, a ground state of $\bar{G}_{U_0}[\Psi]$ never becomes 
a ground state of $\bar{G}_{U}[\Psi]$ with $U\neq U_0$. 
There is a $U$, for which two (or more) states become the ground state 
of $\bar{G}_U[\Psi]$. 
\end{description}
The case I is an irrelevant $U$ case. 
If $|\Psi\rangle$ reproduces $\langle \underline{n}_i^2 \rangle$ 
in a case III, this is an irrelevant $U$ case. 
In the case III or in the case IV, 
we have at least one phase transition with respect to $U$ in the model. 
If $\langle \underline{n}_i^2 \rangle_G$ is reproduced 
by a ground state of a degenerate model in a case IV, 
we have to deny the model, since the degeneracy is 
artificial and the model is improper. 

If we do not have a case II for a Coulomb system, 
although $\bar{F}(U)$ is continuous, 
$\langle\underline{n}_i^2\rangle$ given by 
$|\Psi_U\rangle$ may not be unique because of possible 
degeneracy and may have a jump as a function of $U$. 
The value of $\langle\underline{n}_i^2\rangle_G$ might be 
in this jump. In this case, the model has an artificial phase transition 
and thus the model is not proper. 
Therefor, we have to search for a proper model with 
transferability against change in environment by 
changing $\phi_i$. 
If there exist discontinuity in $\langle\underline{n}_i^2\rangle$ 
along the $U$ axis, 
it is detected by a jump in $\langle\hat{T}\rangle$, 
which is concluded by Lemma \ref{Lemma-statements} ii). 

Finally, we would like to comment on conceptual framework of 
the whole study. You may think that the case II may be special. 
Arbitrariness remains in the selection of $\phi_i$. 
However, we can say that Theorems \ref{Theorem-Uniqueness} 
and \ref{Theorem-Existence} are new guiding principles 
to construct an effective theory of complicated correlated systems. 
In our model, we may choose only one $U$-term, if appropriate. 
This is impossible for a method to make 
an effective model using reduction of the whole phase space by 
choosing a seemingly proper basis. 
Our method is completely different from the renormalization group 
approach. In our method, we keep the property that 
$n_G({\bf r})$ is always reproduced and obtainable. 
Otherwise, an inverse problem, {\it e.g.} the materials design, 
would not be solved properly. 

The author is grateful for valuable comments and suggestions 
on future prospects by Prof. M. Imada and Prof. H. Akai. 
He thanks for fruitful discussion with Prof. N. Suzuki, 
Prof. M. Higuchi, Prof. K. Higuchi and Dr. S. Yamanaka and 
Mr M. Takahashi. The work is partly supported by the 21st COE Program 
by the Japan Society for Promotion of Science and 
a Grand-in-Aid for Scientific Research 
(No. 15GS0213) and (No. 17064006) of the Ministry of Education,  Culture,  
Sports, Science, and Technology, Japan. 


\end{document}